\documentstyle[11pt,IAU207_pasp,twoside,psfig]{article}
\def\edcomment#1{\iffalse\marginpar{\raggedright\sl#1\/}\else\relax\fi}
\reversemarginpar

\begin{document}
\title{New CCD Survey of Globular Clusters in M31}
 \author{Myung Gyoon Lee, Sang Chul Kim}
\affil{Astronomy Program, SEES, Seoul National University, Seoul, 151-742, Korea}
\author{Doug Geisler, Juan Seguel}
\affil{Departamento de F{\'\i}sica, Grupo de Astronom{\'\i}a, Universidad
   de Concepci\'on, Casilla 160-C, Concepci\'on, Chile}
\author{Ata Sarajedini}
\affil{Astronomy Department, University of Florida, Gainesville, FL 32611-2055, USA}
\author{William Harris}
\affil{ Department of Physics and Astronomy,
   McMaster University, Hamilton, ON L8S 4M1, Canada}

\begin{abstract}
We present a progress report of our wide field CCD survey of globular clusters
in M31. 
We have covered 3 deg $\times$ 3 deg area centered on M31, using the KPNO 0.9m
and Washington $CMT_1$ filters.
Our survey is much deeper and more sensitive than previous surveys. 
We have found several hundred new globular cluster candidates in M31
in addition to confirming previously known globular clusters, 
and also have found a number of interlopers among previous globular cluster catalogs.
We have also obtained spectra of about 500 objects among these candidates 
using the HYDRA at the WIYN 3.5m telescope, 
which are used for classification and measuring the radial velocity of the 
candidate objects.
When completed, a new master catalog of globular clusters in M31 will be made,
combining the new globular clusters with the known globular clusters.
\end{abstract}

\section{Introduction}

M31 is an ideal galaxy to study a globular cluster system in a spiral galaxy.
There have been several surveys of globular clusters (GCs) in a large region of M31,
but they were all based on photographic plates. There were a few surveys of 
globular clusters based on CCD observations as well, 
but they were limited to small areas
(see Battistini et al. 1993, Mochejska et al. 1998, Barmby et al. 2000 and references therein).
There are currently $\sim 800$ proposed GC candidates in M31 in the literature.
Over 200 of these objects have been confirmed as GCs, 
200 have been shown not to be clusters, 
and the nature of the remaining objects is as yet unknown (Barmby et al. 2000).
Therefore, over the past few years, we have undertaken a wide field CCD 
survey of the globular clusters in M31.

\section{Observations and Data Reduction}

Using the KPNO 0.9m and the Tek 2k CCD imager, we have obtained
Washington $CMT_1$ images of 53  ($23'\times 23'$) fields covering a region
$3 \times 3$ deg$^2$ centered on M31 to search for new GC candidates.
Figure 1 illustrates a finding chart of M31 showing our survey region.
Then using the WIYN 3.5m and HYDRA (multi-fiber spectrograph) we have obtained
spectra covering {3800\AA--7000\AA}  of about 500 GC candidates.
We have obtained the photometry of the point sources and the extended sources
in the images using DOAPHOT II/ALLFRAME, and have reduced the HYDRA spectra
using the IRAF/DOHYDRA.

\section{Globular Cluster Search Methods}

We have selected globular cluster candidates using several criteria:
a) color-magnitude diagrams, b) color-color diagrams,  
c) morphological classifiers based on the radial moments and the difference
between the aperture magnitude and the point spread function fitting magnitude,
and d) visual inspection of the images.
Our recovery rate of the known bonafide GCs is estimated to be $>90\%$.
Finally we have used the spectra for confirming the GC candidates
(see Seguel et al. 2001 as well).

\section{Results}

Our new photometric survey has produced a total of $\sim 1000$ new GC candidates
with $T_1<20$ ($V<20.5$) mag. About 600 among these candidates are classified
as class 1 (probable GC candidates) and class 2 (possible GC candidates),
 approximately doubling the number of good GC candidates.
Figure 2 shows a color-magnitude diagram of sample new GC candidates as well as
the known GCs, background galaxies and stars in one selected field
(see Kim et al. 2001 for details).
Note how successfully the known GCs are recovered and
how many more new GC candidates (especially faint ones) 
are found in our new survey.
Figure 3 displays $T_1$ luminosity function and  $(C-T_1 )$ color distribution 
of the new GC candidates.
Luminosity function continues to increase with increasing magnitude, passing
the expected turnover at $T_1 \approx 16.7$ mag. This shows 
that the faint end in the luminosity function ($T_1>18$ mag) contains many
non-globular cluster objects.
Color distribution of the GC candidates with $T_1<18-19$ mag shows
a hint of bimodality, but needs a further study to confirm it.
Finally aperture photometry of the GC candidates will be used for 
further analysis.

However, these new GC candidates discovered as part of our photometric survey need
to be confirmed by spectroscopy.
As for the current spectroscopic data of the M31 GCs,
Barmby et al. (2000) contains so far the most comprehensive catalog, 
which still only contains 200 velocity measurements and 
188 spectroscopic metallicities of the GC candidates previously known.
Spectroscopic data are being analyzed now. 
Figure 4 displays sample spectra of new GCs with $T_1=14.9 - 19.0$ mag 
spectroscopically confirmed. Several features typical for globular clusters
(Ca II H and K, H$\beta$,  G band,
Mgb complex, Fe 5270, Na D, and  H$\alpha$ lines) 
are clearly noticed in the spectra.
It is pleasing to find not only new GCs as faint as 19.0 mag 
but also new GCs as bright as 14.9 mag  in our survey.

\section{Future Works}

We are planning to confirm spectroscopically most of the new GC candidates in M31.
When our survey is finished, a new M31 GC catalog will be made, combining
the new GCs and the known GCs.
The final catalog of M31 globular clusters will be used:
1) to  derive a reliable GC luminosity function for M31, 
the faint end of which was very incomplete in  the previous data;
2) to investigate the bimodality of the metallicity distribution, which was
seen only in the spectroscopic data of the previous catalog, 
but not clearly in any photometric data (Barmby et al. 2000); 
3) to investigate the difference in mean luminosity between the metal-poor GCs and
the metal-rich GCs. 
The present catalog of M31 globular clusters shows that the metal-poor GCs are 
0.4 mag fainter than the metal-rich GCs (Barmby et al. 2000, Barmby 2001), 
which is the opposite to the case of the GCs in early-type galaxies and our Galaxy
(Larsen et al. 2001). Our new catalog will be helpful to resolve this discrepancy; and
4) to investigate the systematic difference in kinematics 
between the metal-poor and the metal-rich GCs, which
provides critical information on the age difference between the two
populations and strong constraints on galaxy formation scenarios.


\acknowledgments
This work was supported in part by the Korea Research Foundation Grant (KRF-2000-DP0450).

\begin{figure}
\centerline{
\psfig{figure=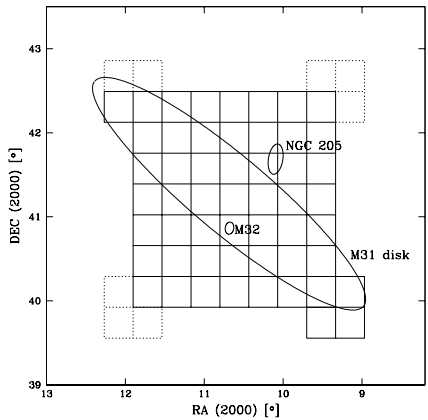,width=6cm,height=6cm}
}
\caption{
A finding chart of M31. 
The solid lines represent the boundaries of our globular cluster survey area,
and the dotted lines represent the boundaries of planned, but not observed area.
}
\end{figure}

\begin{figure}
\centerline{
\psfig{figure=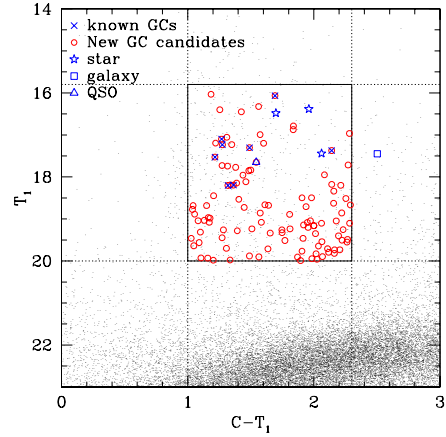,width=6cm}
}
\caption{$T_1$ vs $C-T_1$ color-magnitude diagram of the objects (dots) in 
one selected field in M31. The mass of faint red objects represents
the red giant branch in M31.
The $\times$'s represent previously known globular clusters 
and the $\bigcirc$'s are newly-discovered globular cluster candidates
in seven of our fields in M31.
The square and star symbols represent, respectively, a background galaxy and
stars which were included in previous catalogs of globular
cluster candidates in M31. 
}
\end{figure}

\begin{figure}
\centerline{
\psfig{figure=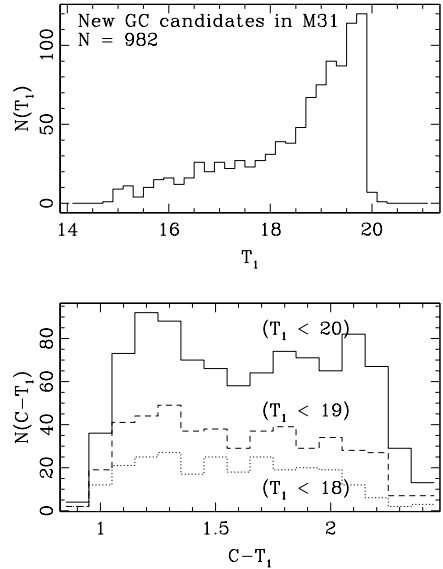,width=6cm}
}
\caption{ (Upper panel) $T_1$ luminosity function of the new globular cluster candidates.
(Lower panel) $(C-T_1 )$ color distribution of the new globular cluster candidates.
}
\end{figure}
\clearpage

\begin{figure}
\centerline{ 
\psfig{figure=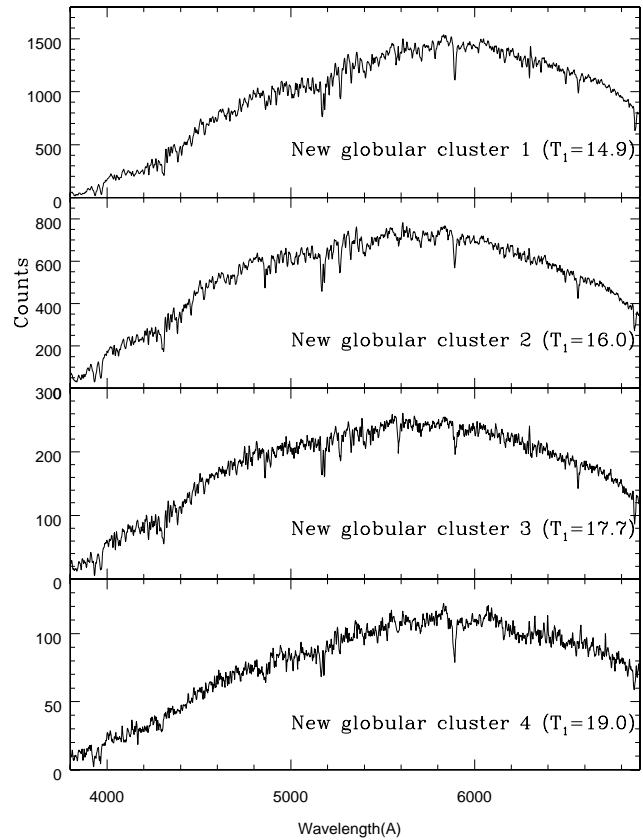,height=12cm} 
}
\caption{Sample spectra of new globular clusters taken with WIYN+HYDRA
in September 2000. 
Note the large range of magnitudes of the new globular clusters,
$T_1=14.9 - 19.0$ mag.} 
\end{figure}

\end{document}